# Hierarchical Architectures in Reservoir Computing Systems

John Moon[1] and Wei D. Lu[1]

[1] Department of Electrical Engineering and Computer Science, University of Michigan, Ann Arbor, MI 48109, United States of America

E-mail: wluee@umich.edu

**Abstract**

Reservoir computing (RC) offers efficient temporal data processing with a low training cost by separating recurrent neural networks into a fixed network with recurrent connections and a trainable linear network. The quality of the fixed network, called reservoir, is the most important factor that determines the performance of the RC system. In this paper, we investigate the influence of the hierarchical reservoir structure on the properties of the reservoir and the performance of the RC system. Analogous to deep neural networks, stacking sub-reservoirs in series is an efficient way to enhance the nonlinearity of data transformation to high-dimensional space and expand the diversity of temporal information captured by the reservoir. These deep reservoir systems offer better performance when compared to simply increasing the size of the reservoir or the number of sub-reservoirs. Low frequency components are mainly captured by the sub-reservoirs in later stage of the deep reservoir structure, similar to observations that more abstract information can be extracted by layers in the late stage of deep neural networks. When the total size of the reservoir is fixed, tradeoff between the number of sub-reservoirs and the size of each sub-reservoir needs to be carefully considered, due to the degraded ability of individual sub-reservoirs at small sizes. Improved performance of the deep reservoir structure alleviates the difficulty of implementing the RC system on hardware systems.

## 1. Introduction

Due to the dramatically increased computing power, advances in algorithms, and abundantly available data, machine learning, especially deep learning [1] has been successfully applied in a wide range of artificial intelligence domains recently, from computer vision to natural language processing. Depending on the tasks, different deep learning architectures have been optimized. For instance, convolutional neural networks (CNNs) [2-4] are mainly used for static data processing such as image recognition, as individual input images are independent so that there is no need to capture the correlation between the input images. On the other hand, temporal data processing such as time-series prediction and speech recognition is typically performed by recurrent neural networks (RNNs) [5-7] since the recurrent connections allow the network to capture temporal features in the sequential input data.

Although the cyclic connections in RNNs are useful to process temporal dynamics of input signals in RNNs, these connections make training RNNs very computationally expensive and difficult. In error backpropagation through time, which is a standard method to train the RNNs, calculating the gradients of the error with respect to weights involves a large amount of computation process as the current states depend on not only the current inputs but also the previous states and inputs. Furthermore, when applying the chain rule to compute the gradients, repeated multiplying the derivatives of the current states with respect to the previous states can cause vanishing gradient or exploding gradient problems [8] that make it difficult to find optimal weights. Long short-term memory (LSTM) [6] is developed to deal with the problems by enforcing constant error flow through internal states of LSTM units. As the weight matrices of the internal gates are trained by gradient-based learning algorithms, LSTM networks have been applied in a broad range of applications by fine-tuning the network for a given task, but nevertheless still requires significant computation costs to train the weight matrices.

Reservoir computing (RC) [7], [9] is proposed to reduce the training cost and mitigate the difficulty of the training process of RNNs. Since the expensive cost and difficulty of training RNNs come from the attempt to control the recurrent connections, RC systems avoid this problem by tuning only the linear output layer (often called the readout layer) instead of training every weight in the recurrent connections of the main RNN body. In RC, the recurrent networks called 'reservoirs' are randomly initialized and fixed during the training process. The reservoir needs to offer fading memory property and should nonlinearly map the temporal inputs into a high-dimensional feature space, represented by the states of the nodes forming the reservoir. Due to the recurrent connections in the reservoir, this nonlinear mapping also involves time. Afterwards, the initial complex inputs can become linearly separable in the new, high dimensional reservoir state space. Software-based RC systems have achieved state-of-the-art performance for tasks such as speech recognition [10] and showed superior ability to forecast large spatiotemporally chaotic systems [11]. Recently, hardware implementations of RC systems have paved the way for applying RC systems to real-time operating hardware systems in power-constrained environments [12-13].

With the successful implementation of RC algorithms and hardware, several methods [14-18] to design the reservoir have been proposed to improve the performance of RC systems. A conventional reservoir is initialized randomly with a single set of hyperparameters such as input scaling and spectral radius, such that the states of the reservoir nodes are coupled strongly and it is difficult to process data with different time scales. Decoupled echo state networks (ESNs) [15] can break the coupling between the nodes by dividing the reservoir into sub-reservoirs and by introducing lateral inhibition unit between the sub-reservoirs. Unlike the conventional random process to build the reservoir, dynamic methods can also be used to generate reservoirs such as the simple cycle reservoir concept [16]. The deterministically constructed reservoirs not only can achieve comparable performance to the conventional randomized reservoir but can also allow reservoir properties to be analyzed in a more comprehensive manner. Additionally, reservoirs using a single physical nonlinear node subjected to delayed feedback [17] have been proposed to allow hardware-friendly implementation of deterministic reservoirs, similar to the simple cycle reservoir concept used in power-efficient hardware systems.

Besides the efforts to improve the single reservoir structure, several studies have been conducted to apply hierarchical structures to RC systems. Early attempts [18-19] to introduce an architectural modification to the reservoir included adding feedforward or static layers in the reservoir to amplify the non-linearity through additional nonlinear mapping of outputs from the recurrent part, and to overcome the tradeoff between short-term memory and non-linearity of a single reservoir structure. With great success of hierarchical architectures in convolutional neural networks [2-4], several works [20-23] have proposed deep reservoir architectures by stacking several sub-reservoirs to capture multiscale dynamics and to increase the richness of reservoir. MESM [20] and DeepESN [21], [22], which stacks the sub-reservoirs directly, and DeePr-ESN [23], which stacks the reservoirs and the encoders alternatively (requiring extra training process), can provide robust prediction performance and effectively capture rich multiscale dynamics. Although the general properties of deep reservoir architectures have been analyzed, emphasis was normally given on the overall performance instead the hardware cost. In reality, simply increasing the size of the reservoirs or the depth of the deep reservoir architecture to achieve better performance may not be optimal given constraints of the hardware system. For example routing the devices acting as reservoir nodes will become increasingly challenging, making it difficult to significantly expand the size of the reservoirs.

In this paper, we show that deep RC systems that stack sub-reservoirs vertically without increasing the overall reservoir size can improve the system performance for a broad range of tasks, by analyzing the distribution of node states in each sub-reservoir and the frequency spectrum captured by each sub-reservoir. To focus on the effects of hierarchical architectures, three different structures based on ESN are compared: Shallow ESN, which has a single reservoir, Wide ESN, which has independent sub-reservoirs in parallel, and Deep ESN, which stacks the sub-reservoirs in series. As the optimal hyperparameters for each structure may be different from each other, a genetic algorithm is used to obtain the best performance of each structure by individually finding the optimal hyperparameters. Moreover, we find that there is a tradeoff between the number of sub-reservoirs and the size of the sub-reservoirs, when the size of the readout network is fixed.

## 2. Architecture

A conventional ESN consists of three basic components: an input layer, a recurrent layer (*i.e.* the reservoir), and an output layer (called the readout layer). The reservoir state, which is the collection of the node states in the reservoir, depends on the input signal and the previous reservoir state. Without losing generality, in our study we choose the following equation to describe the reservoir state update:

$$x(t) = (1 - \alpha) \times x(t-1) + \alpha \times \tanh(W_{in} u(t) + W_{res} x(t-1)) \quad (1)$$

where $u(t) \in \mathbb{R}^{N_U}$, $x(t) \in \mathbb{R}^{N_R}$ denote respectively the input and the reservoir state at time $t$, $W_{in} \in \mathbb{R}^{N_R \times N_U}$ is the input-to-reservoir weight matrix, $W_{res} \in \mathbb{R}^{N_R \times N_R}$ is the recurrent reservoir weight matrix, tanh() is the element-wise applied hyperbolic tangent activation function, and $\alpha \in (0,1]$ is a leaky rate. The weight values in $W_{in}$ is chosen from a uniform distribution over [-$IS$, $IS$], where $IS$ is an input scaling factor. $W_{res}$ is rescaled from the randomly generated $W$, following:

$$W_{res} = SR \times \frac{W}{\lambda_{max}(W)} \quad (2)$$

where $SR$ is the spectral radius, $W$ is chosen from a uniform distribution over [-1,1], and $\lambda_{max}(W)$ is the largest eigenvalue of matrix $W$. $IS$, $SR$, and $\alpha$ are the important hyperparameters affecting the reservoir properties. Typically, $IS$ determines how nonlinear the node states are: a small $IS$ makes the reservoir nodes operate around the origin where the tanh() activation function is virtually linear, while a more nonlinear transformation can be achieved with a large $IS$ that lets the reservoir nodes operate near 1 or -1. $SR$ affects the stability of the node states and should be less than unity to ensure the echo state property, which is a central characteristic of the RC system that states the reservoir state is uniquely defined by the fading history of the input signal. $\alpha$ is related to the speed of the reservoir update dynamics. A small $\alpha$ means the reservoir nodes depend more on the previous reservoir state than on the current input, resulting in a slow update speed. In our approach, these three hyperparameters for each sub-reservoir will be optimized to minimize the error by using the genetic algorithm, as discussed later.

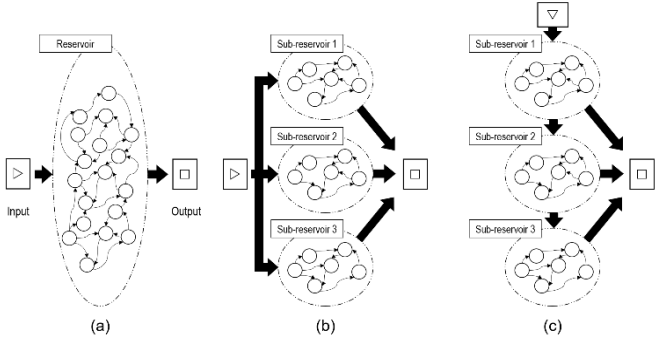

**Figure 1** Reservoir architectures analyzed in this study. (a) Shallow ESN. (b) Wide ESN. (c) Deep ESN.

The output of the ESN is a linear combination of the node states:

$$y(t) = W_{out} x(t) \quad (3)$$

where $y(t) \in \mathbb{R}^{N_Y}$ denotes the output at time $t$, and $W_{out} \in \mathbb{R}^{N_Y \times N_R}$ is the reservoir-to-output weight matrix. In contrast to the fixed weight matrices $W_{in}$ and $W_{res}$ in the reservoir, $W_{out}$ is trained by ridge regression, also known as regression with Tikhonov regularization.

Figure 1 shows the three different ESN structures used to analyze the architectures, namely, Shallow ESN, Wide ESN, and Deep ESN. Shallow ESN is a conventional single reservoir structure, which is governed by (1)-(3). Wide ESN has independent sub-reservoirs in parallel, which receive the input signals independently and have no interaction between the sub-reservoirs. When the sub-reservoirs have different hyperparameters, each sub-reservoir will capture different temporal dynamics from the input signals, and thus Wide ESN may offer better performance than Shallow ESN. The equations of the Wide ESN are following:

$$x^{(l)}(t) = (1 - \alpha^{(l)}) \times x^{(l)}(t-1) + \alpha^{(l)} \times \tanh(W_{in}^{(l)} u(t) + W_{res}^{(l)} x^{(l)}(t-1)) \quad (4)$$

$$y(t) = W_{out}[x^{(1)}(t); x^{(2)}(t); \cdots; x^{(N_L-1)}(t)] \quad (5)$$

where $x^{(l)}(t) \in \mathbb{R}^{N_R}$ denotes the reservoir state of sub-reservoir $l$ at time $t$, $W_{in}^{(l)} \in \mathbb{R}^{N_R \times N_U}$ is the input-to-reservoir weight matrix of sub-reservoir $l$, $W_{res}^{(l)} \in \mathbb{R}^{N_R \times N_R}$ is the recurrent reservoir weight matrix of sub-reservoir $l$, $\alpha^{(l)} \in (0,1]$ is the leaky rate of sub-reservoir $l$, $W_{out}^{(l)} \in \mathbb{R}^{N_Y \times N_L N_R}$ is the reservoir-to-output weight matrix, and $[\cdot; \cdots; \cdot]$ stands for vertical vector concatenation.

Deep ESN is the hierarchical ESN structure that stacks the sub-reservoirs in series. Only the first sub-reservoir can see the input signals, and the subsequent sub-reservoirs receive data from the output of the previous sub-reservoir, in the form of linear combination of the previous sub-reservoir's node states. If stacking the sub-reservoirs improves the system's capability in processing temporal information, Deep ESN will outperform both Shallow ESN and Wide ESN. The equations of Deep ESN are following:

$$x^{(l)}(t) = \begin{cases} (1 - \alpha^{(l)}) \times x^{(l)}(t-1) \\ + \alpha^{(l)} \times \tanh(W_{in}^{(l)} u(t) + W_{res}^{(l)} x^{(l)}(t-1)) & if\ l = 1 \\ (1 - \alpha^{(l)}) \times x^{(l)}(t-1) \\ + \alpha^{(l)} \times \tanh(W_{in}^{(l)} x^{(l-1)}(t) + W_{res}^{(l)} x^{(l)}(t-1)) & if\ l > 1 \end{cases} \quad (6)$$

$$y(t) = W_{out}[x^{(1)}(t); x^{(2)}(t); \cdots; x^{(N_L-1)}(t)] \quad (7)$$

where $x^{(l)}(t) \in \mathbb{R}^{N_R}$ denotes the reservoir state of sub-reservoir $l$ at time $t$, $W_{in}^{(1)} \in \mathbb{R}^{N_R \times N_U}$ is the input-to-reservoir weight matrix for the first sub-reservoir, $W_{in}^{(l)} \in \mathbb{R}^{N_R \times N_R}$ is the inter-reservoirs weight matrix from sub-reservoir $l-1$ to sub-reservoir $l$, $W_{res}^{(l)} \in \mathbb{R}^{N_R \times N_R}$ is the intra-reservoir weight matrix of sub-reservoir $l$, $\alpha^{(l)} \in (0,1]$ is the leaky rate of sub-reservoir $l$, $W_{out}^{(l)} \in \mathbb{R}^{N_Y \times N_L N_R}$ is the reservoir-to-output weight matrix, and $[\cdot; \cdots; \cdot]$ stands for vertical vector concatenation.

The main function of the reservoir in an RC system is to nonlinearly map the sequential input signals onto the reservoir state, which allows certain tasks such as classification and prediction to be processed by the readout network using a linear combination of the reservoir node states. Thus, the quality of the reservoir can be measured in two aspects: how well the reservoir can nonlinearly transform the input signal; and how diverse the temporal dynamics the reservoir can capture. As the performance of the RC system is dominantly affected by how the reservoirs are designed, the hyperparameters for building the reservoirs should be carefully selected to ensure the designed reservoirs produce the best result in given evaluation criteria. To avoid the node states from being easily saturated and ensure the echo state property, $IS$, $SR$, and $\alpha$ are set to be less than 1 as shown in Table I. Grid search is widely used to analyze the influence of hyperparameters on the RC performance, and is a simple way to design the optimal single reservoir system since the hyperparameter space of a single reservoir is not large. However, when the RC system has several sub-reservoirs like Wide ESN and Deep ESN, the grid search is not an efficient optimization method because the computing cost to test every point in the hyperparameter space will increase exponentially as a function of the dimension of the space. Thus, in this paper, a genetic algorithm [23-24], one of the widely used evolutionary optimization methods, is used to find the optimal set of hyperparameters for a given RC architecture.

Genetic algorithms can efficiently explore the large hyperparameter space through the metaheuristic optimization process inspired by biological evolutions. Specifically, a population of sets of hyperparameters for a given RC architecture is randomly generated, and will evolve toward better reservoir structures through an iterative process, where the population with each iteration is called a generation. In each generation, two sets among the population are randomly selected, and two RC systems based on the selected sets are evaluated for a given task by the normalized root mean squared error (NRMSE):

$$NRMSE = \sqrt{\frac{\sum_{t=1}^{T}[y_{target}(t)-y(t)]^2}{\sum_{t=1}^{T}[y_{target}(t)-\overline{y_{target}}]^2}} \quad (8)$$

where $y_{target}(t)$ and $y(t)$ denote the target signal and the output from the RC system at time $t$, $\overline{y_{target}}$ is the mean of the target signal, and $T$ is the number of time steps in the validation sequence. The loser set, which has higher error than the other, winner set, goes through two genetic operations: crossover, which replaces the part of the loser with the part of the winner based on a crossover rate; and mutation, which changes part of the loser to random values based on a mutation rate. The modified loser replaces the original loser in the population, and this process is repeated until the maximum number of generations is reached. The parameters for the genetic algorithm are summarized in Table I.

We evaluate RC systems with different architectures on three different time series data: 10$^{th}$ order nonlinear autoregressive moving average (NARMA10) [25], Santa Fe Laser time series [26], and Mackey-Glass time series [27]. The NARMA10 system is widely used to test ESN models since it needs both nonlinear transformation and memory, which are the main characteristics of RC systems [16-18], [23]. The output of the NARMA10 system is computed as following:

$$y(t+1) = 0.3y(t) + 0.05y(t)(\sum_{i=0}^{9} y(t-i)) + 1.5u(t-9)u(t) + 0.1 \quad (9)$$

where $u(t)$ is a random input at time step $t$, generated from a uniform distribution over [0,0.5], and $y(t)$ is the output at time step $t$. The readout network is trained to predict $y(t+1)$ from the reservoir state and $u(t)$. The lengths of the training, validation, and test sequences are 3000, 100, and 1000, respectively. The first 100 reservoir state values, which are called initial transient, are not used to train the readout network or to predict the next value in the test stage to avoid the influence of initial states. The Santa Fe Laser time series is a one-step ahead prediction on the data obtained by sampling the intensity of a far-infrared laser in a chaotic regime [16], [18]. The data are normalized in the interval [0,1], and the lengths of training, validation, test and initial transient sequences are 3000, 1000, 1000, and 100, respectively. The Mackey-Glass time series is a standard benchmark for chaotic time series prediction task [7], [12], [18], [20], [22]. The time series is defined by the following differential equation:

$$\frac{dy(t)}{dt} = \frac{0.2y(t-\tau)}{1+y(t-\tau)^{10}} - 0.1y(t) \quad (10)$$

where $y(t)$ is the output at time step $t$, and $\tau$ is the time delay. When $\tau > 16.8$, the system has a chaotic attractor. In this study, we set $\tau = 17$, and define the task as to perform 84-step-ahead direct prediction, *i.e.*, the readout network is trained to predict $y(t + 84)$ from the reservoir state when $y(t)$ is given. The lengths of training, validation, test and initial transient sequences are 1000, 1000, 1000, and 100, respectively.

TABLE I
HYPERPARAMETERS FOR RESERVOIR AND
PARAMETERS FOR THE GENETIC ALGORITHM

| Hyperparameters for Reservoir | Symbol | Value |
|---|---|---|
| Input scaling | IS | (0, 1] |
| Spectral radius | SR | (0, 1] |
| Leaky rate | $\alpha$ | (0, 1] |

| Parameters for the Genetic Algorithm | Value |
|---|---|
| Generation | 1000 |
| Population size | 15 |
| Crossover rate | 0.33 |
| Mutation rate | 0.33 |

In the genetic algorithm, the NRMSE is used to determine which hyperparameter set is the winner, as shown in Figure 2(a). Additionally, the distribution and frequency spectrum of the node states are measured to study the properties of different reservoir structures and to explain why a specific reservoir structure performs better than others. To get better performance, the RC system should nonlinearly transform the input signals into a high-dimensional space and allow the readout network to linearly separate the corresponding reservoir states. Although it is easy to decide whether the transformation is linear or nonlinear, it is difficult to quantify the degree of nonlinearity in a single evaluation term. Here, the distribution of node states is analyzed since nodes whose states are far away from the origin indicate these nodes go through a more nonlinear transformation than the nodes whose states are close to the origin. After finding the optimal hyperparameter set for a given reservoir structure and task, the node states of the test sequence are recorded, and the mean and standard deviation of the node states are calculated as shown in Figure 2(a) To clearly visualize the difference of each sub-reservoir, the node states are grouped according to the sub-reservoir they belong to, and the means of node states in the same sub-reservoir are sorted in ascending order.

Another important property of the RC system is the ability to capture diverse temporal dynamics with different time scales. The diversity of temporal dynamics can be analyzed by examining the differences of the frequency components each sub-reservoir captures. The multiple superimposed oscillator (MSO) task [15], [28-29] is used to analyze the multiple time-scales processing ability of the different reservoir structures. The MSO12 time series, given by a sum of sinusoidal functions, is used:

$u(t) = \sum_{i=1}^{12} \sin(\varphi_i t)$ (11)

where $\varphi_i$ determines the frequency of the i-th sinusoidal function. The $\varphi_i$ coefficients are set as in [28], *i.e.* $\varphi_1 = 0.2, \varphi_2 = 0.331, \varphi_3 = 0.42, \varphi_4 = 0.51, \varphi_5 = 0.63, \varphi_6 = 0.74, \varphi_7 = 0.85, \varphi_8 = 0.97, \varphi_9 = 1.08, \varphi_{10} = 1.19, \varphi_{11} = 1.27, \varphi_{12} = 1.32$. As shown in Figure 2(b), the frequency spectra of each node state after being excited by the MSO12 input are calculated by the Fast Fourier Transformation (FFT). The FFT values are averaged on a sub-reservoir-by-sub-reservoir basis. To emphasize which frequency components are mainly captured by the specific sub-reservoir, the peak values of the 12 different frequency components are normalized by the minimum among them.

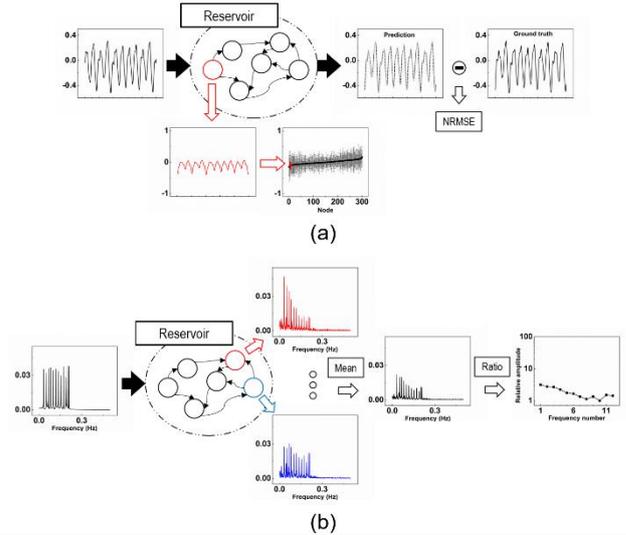

**Figure 2** Schematic of the evaluation methods. (a) NRMSE and the distribution of node states. NRMSE is measured from the difference between the prediction of the test sequence and the ground truth. The temporal response (red line) of a node (*e.g.* represented by the red circle) is recorded for a test sequence. The mean and standard variation of the states from each node in a sub-reservoir are plotted and sorted in ascending order by the mean value. (b) The frequency spectrum of the node states. When the MSO12 time series is applied to the reservoir, the FFT of the states from each node is calculated (*e.g.* red, blue lines) and averaged (black line) on a sub-reservoir-by-sub-reservoir basis. The peak values of the 12 different frequency components are normalized by the minimum among them.

## 3. Results

In this section, we will compare the performance of the three different reservoir structures on time series prediction tasks, and discuss the effects of stacking sub-reservoirs on the properties of the RC system.

As the performance of an RC system is strongly affected by the size of the readout network, we first fixed the size of the readout network to 300, *i.e.* corresponding to a single reservoir with 300 nodes (Shallow ESN), three independent sub-reservoirs with 100 nodes each (Wide ESN), and three stacked sub-reservoirs with 100 nodes each (Deep ESN). Figure 3(a), (b), and (c) show the NRMSEs of the three different reservoir structures when performing NARMA10, Santa Fe Laser and Mackey-Glass time series tasks, respectively. The box charts show the results from 10 different randomly generated reservoir systems with the hyperparameter set optimized by the genetic algorithm. In all cases, Deep ESN shows the best performance among the different reservoir structures. In contrast, compared to Shallow ESN, Wide ESN shows slightly better performance for the first two tasks but worse performance for Mackey-Glass time series.

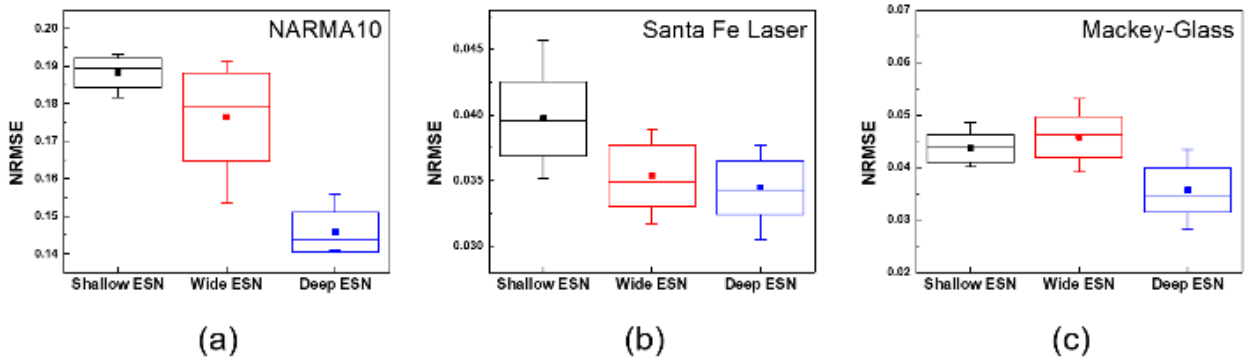

**Figure 3** NRMSE of Shallow ESN, Wide ESN and Deep ESN for (a) NARMA10 system. (b) Santa Fe Laser time series, and (c) Mackey-Glass time series tasks. The boxes show the 25th, 50th and 75th percentiles for the NRMSE data measured from 10 random reservoir initializations for each case, along with the mean (dot). The whiskers represent the minimum and maximum.

To find out why Deep ESN performs better than others, the distributions of node states are plotted in Figure 4. The nodes of Deep ESN in the third (*i.e.* last) sub-reservoir show a wider spectrum than not only the nodes of Deep ESN in the first two sub-reservoirs, but also the nodes in every sub-reservoir in other tested structures. This indicates that the nodes of the third sub-reservoir in the Deep ESN go through a higher degree of nonlinear transformation. Considering the nodes of Deep ESN in the deeper sub-reservoir have a wide spectrum than the nodes in the previous sub-reservoirs, the wide range of node states in Deep ESN can be attributed to the stacked sub-reservoir structure in the Deep ESN.

Besides the absolute magnitude of the node states range, the diversity of the ranges among the sub-reservoirs is also a good factor to examine the quality of the reservoir structure, because having different ranges between the sub-reservoirs means that the node states in each sub-reservoir have different degrees of nonlinear transformation so that the RC system can capture broader temporal dynamics. It can be seen from Figure 4 that Wide ESN has a slightly larger range and diversity of node states than Shallow ESN, but the enhancement is not as significant as Deep ESN. Thus, a more efficient way to achieve a high degree of diverse, nonlinear transformation is stacking the sub-reservoirs in series rather than building independent sub-reservoirs in parallel.

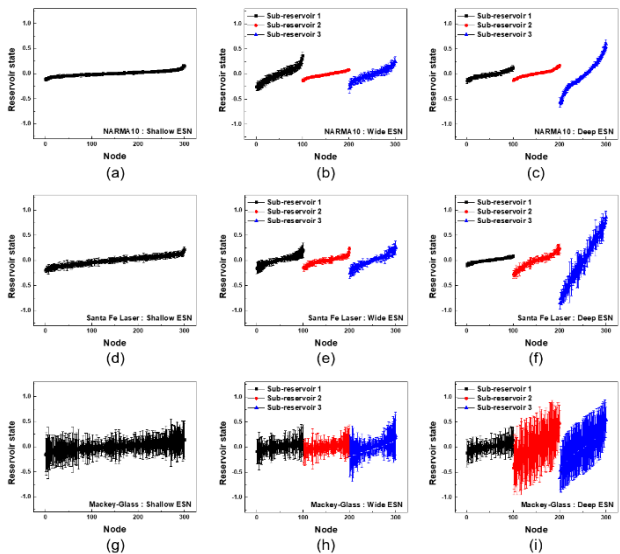

**Figure 4** Distribution of the mean and standard deviation of the node states. (a) Shallow ESN, (b) Wide ESN, (c) Deep ESN for the NARMA10 system. (d) Shallow ESN, (e) Wide ESN, (f) Deep ESN for Santa Fe Laser time series. (g) Shallow ESN, (h) Wide ESN, (i) Deep ESN for Mackey-Glass time series. The node states in the same sub-reservoir are sorted in ascending order of their mean values.

Another method to estimate the temporal dynamics captured by each sub-reservoir is the frequency spectrum of node states, which shows which frequency components are picked up by each sub-reservoir. In our study, due to the simple but distinguishable frequency spectrum of the MSO12 input signal, it makes it easier to recognize the distinct frequency spectra among the reservoir structures, as shown in Figure 5. While Wide ESN shows similar frequency spectra among the sub-reservoirs, the sub-reservoirs in Deep ESN show significantly different frequency spectra. Specifically, the sub-reservoirs in the late stages of Deep ESN tends to capture the low frequency component rather than the high frequency component. Physically, this can be attributed to the fading memory property of the (sub-)reservoir which essentially performs an temporal integration function of the input sequence. The resulting reservoir state, reflecting the integrated input, is then used as inputs to the next stage, allowing lower frequency components from the original input to be more efficiently captured in the later stages of Deep ESN. Analogous to the fact that the layers in the late stage of CNNs capture more abstract or global information, *i.e.* lower spatial frequency features of the input signal, the sub-reservoirs in the late stages of Deep ESN thus emphasizes on the low temporal frequency component or the global (long term) temporal information. We also note that, although both CNNs and Deep ESNs put similar roles to the layers in the late stage, they utilize the early stage layers significantly differently. Typically a CNN only uses the outcome of the last layer since its goal is to derive an abstract representation of the input signal, such as identifying the object, and the outcome of the last layer is typically enough for the task. On the other hand, Deep ESN needs to utilize the reservoir states of every sub-reservoir, since a wide range of temporal information from low to high frequency components is usually required to perform the temporal data processing tasks such as predicting the next value of the time series data.

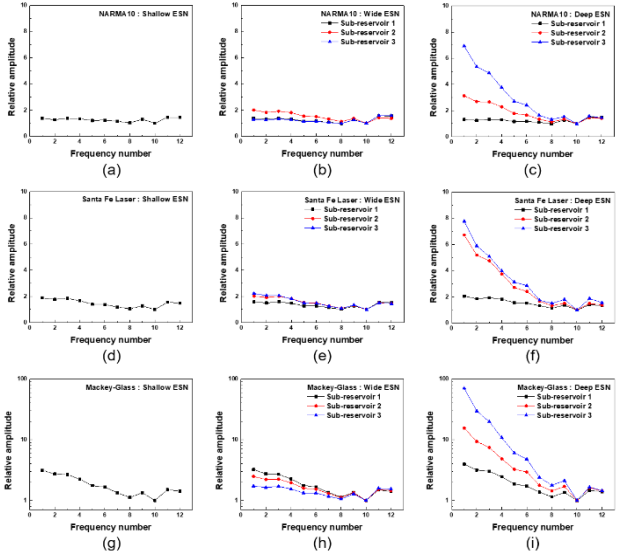

**Figure 5** Frequency spectrum of node states. (a) Shallow ESN, (b) Wide ESN, (c) Deep ESN for the NARMA10 system. (d) Shallow ESN, (e) Wide ESN, (f) Deep ESN for Santa Fe Laser time series. (g) Shallow ESN, (h) Wide ESN, (i) Deep ESN for Mackey-Glass time series.

The simplest way to improve the performance of the RC system is increasing the total size of the reservoir, since larger reservoirs with more diverse internal connections should capture richer temporal dynamics. If stacking the sub-reservoirs in series helps the RC system to predict the time series data more accurately, Deep ESN should still outperform the others when the total size of the reservoir increases. To expand the size of the reservoir, extra sub-reservoirs with 100 nodes are serially added to the output of the last sub-reservoir in Deep ESN, or connected as parallel, independent sub-reservoirs in Wide ESN. Shallow ESN has still a single reservoir, but the size of the reservoir increases to match the total size of the expanded Deep ESN or Wide ESN. Figure 6(a), (b) and (c) show the NRMSEs of the three different reservoir structures on NARMA10, Santa Fe Laser and Mackey-Glass time series, respectively, as the total size of the reservoir increases from 200 to 500. Again Deep ESN shows the lowest NRMSE among the tests, and Wide ESN is better than Shallow ESN in most cases, which are consistent with the findings for the case with 300 total nodes. In general, as the total number of nodes in the reservoir increases, the performance of the RC system improves no matter what reservoir structure is used, but the improvement in prediction error tends to be saturated. Moreover, when considering both the performance measured in NRMSE and the computation cost measured in the size of weight matrices, the performance improvement by simply increasing the reservoir size may actually be lower than the increased expense needed to compute the reservoir states and train the readout network.

Instead of adding extra sub-reservoirs to Wide ESN or Deep ESN, we can instead fix the total size of the reservoir (thus keeping the total compute cost low) and vary the number and size of sub-reservoirs. While Shallow ESN always has a single reservoir with 300 nodes, we varied the number of sub-reservoirs in Wide ESN and Deep ESN from 2 to 5 while keeping the total number of nodes constant at 300. Due to the fixed total node size, the cost to train the readout network will be identical for the three different reservoir structures, but the expense to calculate the node states will be reduced as the weight matrix size in Wide ESN or Deep ESN is reduced quadratically as the sub-reservoir size is reduced, while the number of matrices increases linearly. Figure 7(a), (b) and (c) show the NRMSEs of the three different reservoir structures on NARMA10, Santa Fe Laser and Mackey-Glass time series tasks, respectively, as the number of sub-reservoirs increases from 2 to 5 while keeping the total node size at 300. Again, Deep ESN outperforms the others for all tasks in all cases.

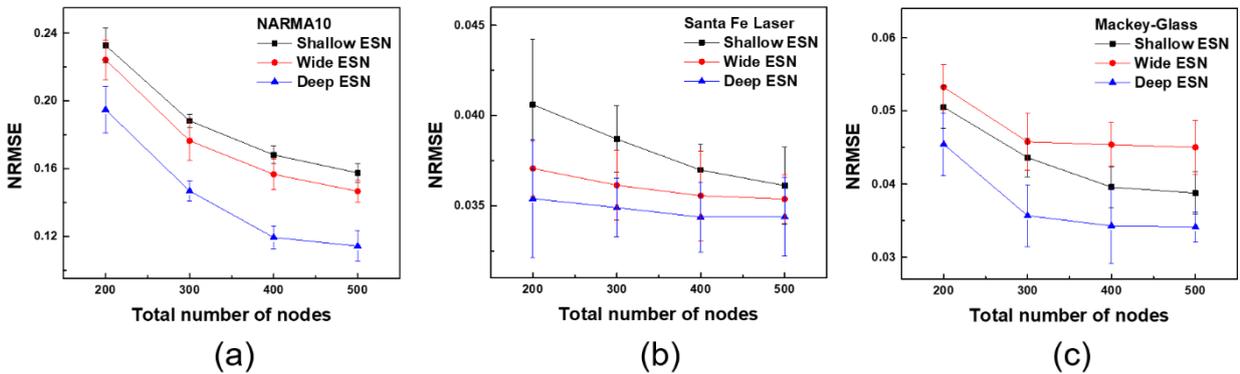

**Figure 6** NRMSE of Shallow ESN, Wide ESN and Deep ESN with different total numbers of nodes in the system, for (a) NARMA10 system, (b) Santa Fe Laser time series, and (c) Mackey-Glass time series. The mean and standard variations are plotted for each case, based on 10 different random initializations of the reservoirs following the optimized hyperparameter sets.

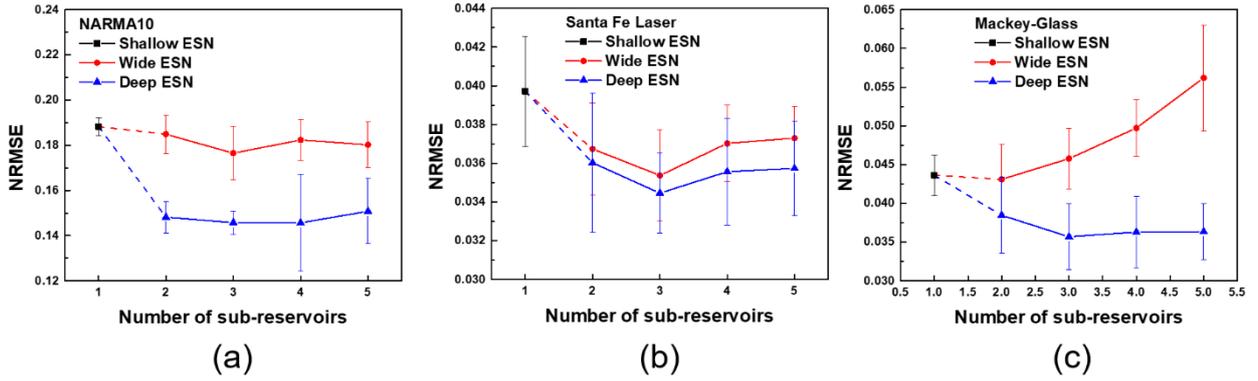

**Figure 7** NRMSE for Shallow ESN, Wide ESN and Deep ESN with different number of sub-reservoirs, for (a) NARMA10 system, (b) Santa Fe Laser time series, and (c) Mackey-Glass time series. The mean and standard variations are plotted for each case, based on 10 different random initializations of the reservoirs following the optimized hyperparameter sets.

For the NARMA10 and Santa Fe Laser tasks, Wide ESN and Deep ESN show similar trend on the number of sub-reservoirs, *i.e.* the performance is improved when the reservoir consists of 2 or 3 sub-reservoirs, but becomes worse when the reservoir is divided into more than 2 or 3 sub-reservoirs. This can be explained by the tradeoff between the number of sub-reservoirs and the size of each sub-reservoir. Given the fixed total number of nodes, the size of sub-reservoirs is reduced as the number of sub-reservoirs increases. When the reservoir is split into several sub-reservoirs, the temporal dynamics captured by the whole system become more diverse since each sub-reservoir can have different hyperparameters leading to different temporal dynamics. Moreover, stacking the sub-reservoirs enhances the nonlinearity of transformation from the input signal and the diversity of the temporal dynamics captured by each sub-reservoir, as shown in Figure 4 and 5. However, at the same time, the feature space of each sub-reservoir will shrink due to the reduced size of the sub-reservoirs, *i.e.* the ability of the sub-reservoirs to extract the temporal information will be weakened.

The optimal number and size of the sub-reservoirs depend on both the reservoir structure and the task. For example, the result for Mackey-Glass time series shown in Figure 7(c) is different from the trends observed in Figure 7(a) and (b). As the number of sub-reservoirs increases, Deep ESN performance always improves, but Wide ESN performance becomes worse. This clear dependence on tasks originates from the intrinsic characteristics of each task. Figure 8(a), (b) and (c) show the FFT results of the input signals of NARMA10, Santa Fe Laser and Mackey-Glass time series, respectively. While the output of NARMA10 is governed by the input signal based on (9), the input of NARMA10 is randomly generated, not affected by the output. Thus, the frequency spectrum of the input signal of NARMA10 has no major frequency dependence. In contrast, Santa Fe Laser and Mackey-Glass time series have clear frequency peaks since in these tasks the input is the previous output, not an independent variable. Importantly, the major peaks of Mackey-Glass time series are located near the low frequency range rather than spreading out evenly, which is the case of Santa Fe Laser time series. As shown in Figure 8(d), (e) and (f), the sub-reservoirs in the late stage of Deep ESN can effectively capture the low frequency components. Especially for Mackey-Glass time series, this strength of Deep ESN can compensate the degradation of the sub-reservoir feature space. Thus, the NRMSE of Deep ESN for Mackey-Glass time series improves as the number of sub-reservoirs increases. On the other hand,

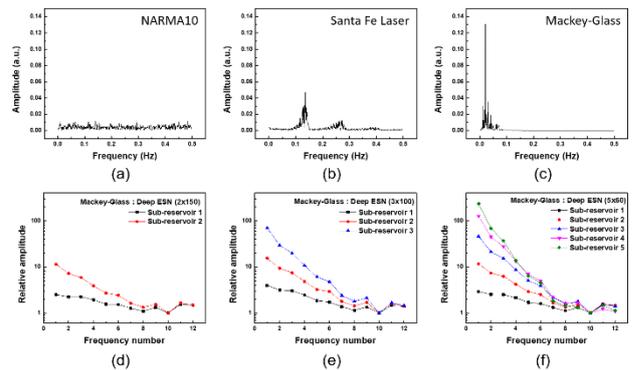

**Figure 8** Fast Fourier Transform results of the input signal of (a) NARMA10, (b) Santa Fe Laser time series, and (c) Mackey-Glass time series, and frequency spectra of node states of Deep ESN with 2 sub-reservoirs having 150 nodes each (d), 3 sub-reservoirs having 100 nodes each (e), and 5 sub-reservoirs having 60 nodes each (f), for the Mackey-Glass time series problem.

Wide ESN performance for Mackey-Glass time series becomes worse when having more sub-reservoirs.

In the cases of NARMA10 and Santa Fe Laser time series, the improvement comes from having more diverse temporal dynamics rather than extracting more information from the low frequency component, as both tasks do not have major peaks in the low frequency range. Thus, although the Deep ESN still outperforms Shallow ESN and Wide ESN, the benefits in capturing more low-frequency signals are not fully utilized and the Deep ESN performance degrades when the sub-reservoir size becomes too small. Following similar reasoning, even small differences among the independent sub-reservoirs of Wide ESN contribute to the improvement of Wide ESN over Shallow ESN for these two tasks.

Finally, we compare our Deep ESN with other networks. For the task of predicting Mackey-Glass time series 84 step ahead, the Deep ESN (NRMSE = 0.034) outperforms reported performance from feed-forward neural networks (NRMSE = 0.038) [30] and LSTM (NRMSE = 0.470) [31] because of the Deep ESN's ability to capture diverse temporal dynamics. The small number of units and trainable parameters (4 units and 113 trainable parameters, respectively) may contribute to the worse reported LSTM performance than the feed-forward neural networks. However, although the performance of LSTM network can be improved by increasing the number of units, the training cost will also increase dramatically because the weight matrix size of internal gates increases quadratically. Moreover, the gradient-based learning algorithm is computationally more expensive than ridge-regression, which is the learning rule commonly used for RC systems. Among reported methods to design RC systems, DeePr-ESN (NRMSE = 9.08E-04) [23] shows better performance than the Deep ESN analyzed here, but DeePr-ESN inserts additional encoder layers between sub-reservoirs, which requires extra training for auto-encoders based on extreme learning machines. The large total size of DeePr-ESN (8 sub-reservoirs with 300 nodes each vs. 5 sub-reservoirs with 60 nodes each for the Deep ESN) is another factor of the better DeePr-ESN performance.

## 4. Discussion

Hierarchical reservoir structures are investigated in this study to understand how the different reservoir architectures affect the performance of the RC system. While Shallow ESN can pick up only limited temporal dynamics from the time-series input due to a single set of hyperparameters, Wide ESN and Deep ESN can capture more diverse temporal dynamics by designing each sub-reservoir with different sets of hyperparameters. The enhancement of Wide ESN is not as significant as Deep ESN since the genetic algorithm does not have any external momentum to assign different hyperparameters to the sub-reservoirs of Wide ESN. Other hyperparameter optimization methods that force each sub-reservoir to have distinct hyperparameters may be helpful to improve the Wide ESN performance. In addition to the enhanced diversity in temporal dynamics, Deep ESN provides higher degree of nonlinear transformation, as well as the ability to efficiently capture the low frequency components. The genetic algorithm, which already led to good optimization results in DeePr-ESN [23], allows the large hyperparameter space in the hierarchical structures to be explored with reasonable computational cost, so that optimized hyperparameter sets for Deep ESN can be obtained.

In addition to the nonlinearity and diversity of reservoir transformation, the memory property of a reservoir structure can be evaluated by the memory capacity (MC), which measures the ability of the reservoir to store and recall previous inputs by computing correlations between the delayed inputs and the reconstructed outputs [32-35]. MC can be affected by the reservoirs' hyperparameters such as input scaling and spectral radius. It was found that the maximum MC can be obtained with a linear reservoir in the absence of numerical errors [33]. Although an ideal reservoir should have a high degree of nonlinearity and a large MC, there is a trade-off between the nonlinearity and MC [34-35] because reconstructing the delayed input is a kind of linear mapping from the original input to the delayed input, which conflicts with the nonlinearity property . Similarly, we observed the tradeoff between nonlinearity and MC in this study. For instance, Deep ESN shows a higher degree of nonlinearity compared to Wide ESN and Shallow ESN in the NARMA10 task, as shown in Figure 4, but produces a smaller MC (20.17±0.83) than Wide ESN (21.35±0.58) and Shallow ESN (23.64±0.47). However, the large hyperparameter space offered by the hierarchical structure means Deep ESNs can also provide large MC by optimizing the hyperparameters, where MC larger than Wide ESN and Shallow ESN has been reported in previous studies [21]. In other words, depending on which property (*e.g.* nonlinearity, MC) is more critical for a specific task, the desired property can be enhanced in Deep ESN with proper hyperparameter optimization.

While software-based reservoir computing systems have broad design options such as aggressively expanding the reservoir size and inserting encoder layers between sub-reservoirs, several constraints have to be considered when designing hardware-based reservoir computing systems. For instance, physically connecting the nodes is not trivial because the complexity of routing large number of devices grows exponentially when the size of reservoir increases. Moreover, if the device is passive, active components that control the signal flows from one device to others should be also carefully designed. Due to these physical constraints, hardware-based reservoir computing systems have not shown as fast improvement in performance as software-based reservoir computing systems, even though several studies have demonstrated promising features of hardware-based reservoir computing systems such as power-efficiency and computing speed [13]. The hierarchical reservoir structures studied here respect the hardware constraints and achieves better performance by capturing more diverse temporal dynamics and enhancing the degree of nonlinear transformation without increasing the reservoir size. In fact, the number of interconnections between devices is actually reduced due to the use of sub-reservoirs to alleviate the complexity of routing. Furthermore, when the total reservoir size is fixed, the analysis of the tradeoff between the number of sub-reservoirs and the size of each sub-reservoir allows one to design an optimal hardware architecture by considering the balance between the effects of increased hierarchical depth and the capacity sub-reservoirs.

## 5. Conclusion

In this paper, the effects of hierarchical reservoir structures on the performance of reservoir computing are investigated. Wide-ESNs that build independent sub-reservoirs in parallel, and Deep-ESNs that stack the sub-reservoirs in series, can capture more diverse temporal dynamics than simply increase the reservoir size. Furthermore, Deep-ESNs not only offers stronger nonlinear transformation of the input to the reservoir states, but also captures more diverse temporal dynamics than Wide-ESNs and Shallow-ESNs. When the total size of the reservoir is fixed, a trade-off may be observed between the number of sub-reservoirs and the size of each sub-reservoir. For tasks where the low frequency signals are important, such as long-term forecasting of time-series such as Mackey-Glass time series, the strength of Deep ESN becomes more evident as the enhanced ability of the sub-reservoirs in the late stage to capture low frequency signals overcomes the effect of reduced feature space of the sub-reservoirs.

The analysis of hierarchical reservoir structures may provide useful insight into designing practical reservoir systems. Especially, when it comes to hardware implementation of RC systems, reducing the number of connections between the nodes by stacking sub-reservoirs offers a promising approach to building large reservoirs in hardware. Enhancement of nonlinearity and diversity from the stacked sub-reservoir structure should be considered and utilized for solving complex tasks such as multivariate systems. Further studies on hyperparameter optimization algorithms can help fine tune the reservoir design for specific tasks, and provide a more comprehensive understanding of how the specific hyperparameter set affects the overall performance of the RC system.

## Acknowledgements

The authors thank Dr. M. A. Zidan for helpful discussions. This work was supported in part by the National Science Foundation through awards CCF-1900675.